\documentstyle[prb,aps,multicol]{revtex} \tighten \begin{document} \draft
\title{Screening of a macroion by multivalent ions: A new boundary
condition for the Poisson-Boltzmann equation and charge inversion.}
\author
{V. I. Perel~\cite{Perel} and B. I. Shklovskii} \address{Theoretical Physics
Institute, University of Minnesota, 116 Church St. Southeast, Minneapolis,
Minnesota 55455} \maketitle \begin{abstract}
 Screening of a macroion by
multivalent counterions is considered.
It is shown that ions form strongly correlated liquid at the macroion surface.
Cohesive energy of this liquid leads to strong additional attraction of counterions 
to the surface. Away from the
surface this attraction is taken into account by a new boundary
condition for the Poisson-Boltzmann equation. 
This equation is solved with the new boundary condition for a charged
flat surface and a long cylinder. For a cylinder Onsager-Manning theory looses
its universality so that apparent charge of the cylinder
is smaller than their theory predicts and depends on its bare charge. It
can also vanish or change sign.
\end{abstract}

 \begin{multicols}{2}

\section {Introduction}

Many polymers are strongly charged in a water solution. Lipid membranes,
DNA and actin
are the most famous examples of such biological polyelectrolytes. In a
water solution polyelectrolytes are
 screened by smaller ions of both signs.
Correct description of the screening of polyelectrolytes is tremendously
important for calculation of properties of individual polyelectrolytes, for
example, the effective charge or the
bending rigidity. The screening also determines forces acting 
between polyelectrolytes and both thermodynamic
and transport properties of their solutions. Here we concentrate on a rigid
polyelectrolyte with a fixed charge distribution. Two standard problems
are considered below $-$ an infinite flat surface and an infinite cylinder $-$
each uniformly charged with the surface density
$- \sigma <0$. The standard approach
for a description of such problems is the Poisson-Boltzmann
equation (PBE) for the selfconsistent electrostatic potential $\psi({\bf r})$
\begin{equation}
\nabla^{2} \psi = -{{4 \pi e}\over{D}} \sum Z_{i}N_{0i}
\exp\left(-{{Z_{i}e\psi}\over{k_BT}}\right).
 \label{PB}
\end{equation}
Here $e$ is the charge of a proton, $D \simeq 80$ is the dielectric constant of
water, $Z_{i}e$ is the charge
of a small ion of sort $i$ and $N_{0i}$ is their concentration at the
point where $\psi=0$.
 The number of papers using the analytical and numerical solutions of
Eq.~(\ref{PB})
 is extremely large~\cite{Frank}.
On the other hand there is understanding that Eq.~(\ref{PB}) neglects
ion-ion correlations and is not exact.
Deviations from the distribution of charge predicted by PBE were
demonstrated numerically~\cite{Gulbrand,Roland}
 for the following problem. Consider screening of a charged surface, $x=0$,
of a membrane or a film by a water solution occupying halfspace $x>0$.
Assume that there is only one sort of counterions with charge $Ze > 0$
and their concentration $N(x) \rightarrow 0$
at $x \rightarrow \infty$. In this case solution of Eq.~(\ref{PB})
is very simple and has the Gouy-Chapman form
\begin{equation}
\label{GC}
N(x) = {1\over{2 \pi l}}{1\over {(\lambda + x)^2}},\\ \end{equation}
where $\lambda = Ze/(2 \pi l \sigma)$ is the Gouy-Chapman length, $l=Z^2 l_{B}$
and $l_{B} = e^{2}/(Dk_B T) \simeq 0.7$ nm is the Bjerrum length.
At large $Z$ and $\sigma$, the length $\lambda$ can become
of the order of the size of the water molecule or even smaller.
For example, at $Z=3$ and $\sigma = 1.0~e/$nm$^{2}$ length $\lambda = 0.08$ nm.
Thus, almost all counterions are located in the first molecular layer at
the surface or, in other words, they condense at the very surface of
polyelectrolyte.
This raises question about a role of their lateral correlations and validity
of Eq.~(\ref{GC}).
It was found by numerical methods~\cite{Roland} that for a typical charge
density $\sigma$ deviations from Eq.~(\ref{GC})
are not large for monovalent
counterions, but they strongly increase with the charge of counterions $Z$.
It was suggested in
Ref.~\onlinecite{Rouzina96,Bruinsma,Levin,Shklov98}
that at $Z \geq 2$
repulsion between multivalent counterions condensed at the surface
 is so strong that they form a two-dimensional strongly correlated liquid (SCL)
 in which the short order of counterions is similar to that of Wigner crystal (WC).
This idea was used to demonstrate that two charged surfaces in the presence of
 multivalent counterions can attract each other at small distances.

The goal of this paper is to develop a simple analytical theory of the
influence of SCL on a screening
atmosphere of a single charged surface. It is shown below that the cohesive
energy of SCL provides additional binding
of ions to the surface. PBE fails to describe this correlation effect. At the same time,
PBE works well far from the surface where
$N(x)$ is small  and correlations are not important.
 To describe screening at large distances we derive a new boundary
condition for $N(x)$ at $x=0$, which takes the effect of SCL
into account. Then we solve PBE with this boundary condition for standard
problems of
screening of charged flat surface and a cylinder for different salt
concentration $N(\infty)$ in the bulk of solution.
In the case of a cylinder, we show that
in practical conditions the conventional picture
of the Onsager-Manning~\cite{Manning}
condensation should be strongly modified when dealing with multivalent ions.
Since the counterions are tighly bound to the surface of cylinder the
net linear charge density
of a screened cylinder is smaller than in the Onsager-Manning theory and 
depends on a bare density. The net charge density can also cross zero and
become positive.
In the case of a flat surface similar phenomena are found which lead
to the screening atmosphere different from  Eq.~(\ref{GC}) and are in
qualitative agreement with numerical results~\cite{Roland}.
Because of simplicity and the use of the universal boundary condition our
theory complements direct numerical methods~\cite{Roland} which until now were not
able to study the case of $Z\geq 3$ and other than planar geometries.

Let us present arguments for the existence of SCL and then derive a new
boundary condition for $N(x)$.
As we mentioned above and will show below for a typical
$\sigma \geq 1 e$ nm$^{-2}$ and $Z \geq 3$, almost all charge of the plane
is compensated by SCL of counterions at its surface, which has a
two-dimensional concentration almost equal to
$ n = \sigma / Ze$.
 The minimum of Coulomb energy of
counterion repulsion and their attraction to the background is provided by
triangular close packed WC of
counterions~\cite{mara}. At $T=0$ the energy per ion of such WC,
$\varepsilon(n)$,
can be  estimated as the energy of attraction, $-Z^{2}e^{2}/DR$, of an
counterion to its Wigner-Seitz cell
(correlation hole) which is approximately a disc with radius
$R = (\pi n)^{-1/2}$ and charge $-Ze$. More accurately
\begin{equation}
\varepsilon(n)= - \alpha n^{1/2}Z^{2}e^{2}D^{-1},
\label{energyion}
\end{equation}
where $\alpha=1.96$.
The inverse dimensionless temperature of SCL is usually written in units
\begin{equation}
\Gamma = \frac{Z^{2}e^{2}}{R D k_BT} = 0.9\frac{|\varepsilon(n)|}{k_BT}.
 \label{Gamma}
\end{equation}
For example, at $\sigma = 1.0~e/$nm$^{2}$ and room temperature,
Eq.~(\ref{Gamma}) gives $\Gamma =3.5,~6.4,~9.9$
at $Z=2,~3,~4$. Thus for multivalent counterions we
deal with low temperature situation. $\Gamma$
is the large parameter of our theory. 
In its terms $R /\lambda \simeq 2\Gamma \gg 1$
 and $l/R \simeq \Gamma \gg 1$.
For example, at $Z=3$ and $\sigma = 1.0~e/$nm$^{2}$ lengths $\lambda,
R$ and $l$ are equal to 0.08,~1.0,~6.3 nm
respectively.

It is known~\cite{Gann} , however, that  WC melts at very low
temperature near $\Gamma \simeq 130$. 
So in the range of our interest, $3 < \Gamma < 15$,
we deal with SCL. Thermodynamic properties of such SCL or one
component plasma were studied numerically
~\cite{Totsuji,Gann}. In the large range
 $0.5 < \Gamma < 50$ excessive internal energy of SCL per counterion,
$\varepsilon(n,T) = k_BT f(\Gamma)$,
was fitted by the expression~\cite{Totsuji}
\begin{equation}
f(\Gamma) = - 1.1 \Gamma + 0.58 \Gamma^{1/4} - 0.26
\label{intenergy}
\end{equation}
with an error less than 2\%. The first term on the right side of
Eq.~(\ref{intenergy}) dominates at large $\Gamma$ and
leads to Eq.~(\ref{energyion}).
The other two terms provide a relatively small correction to the 
energy of WC. It is equal to 11\% at $\Gamma = 5$ and to 5\% at $\Gamma = 15$. 
The reason for a such small difference is that short 
range order in SCL is similar to that of WC.
For the free energy per unit area, $F$, we can write
$F=F(\Gamma=0.5) + nk_BT\int_{0.5}^{\Gamma}f(\Gamma^{\prime}){d\Gamma^{\prime}}/\Gamma^{\prime}$,
so that for the chemical potential which we need below to describe the
equilibrium of SCL with the gas-like phase we obtain
\begin{eqnarray}
\mu(n,T) &=& - k_BT\ln (n_w/n) + \mu_{s} + \mu_c(n,T),
 \label{sumchem} \\
\mu_c(n,T) &=& - k_BT (1.65 \Gamma
- 2.61 \Gamma^{1/4} + 0.26 \ln\Gamma + 0.13). \label{cpot}
\end{eqnarray}
Here $\mu_c$ is contribution of correlations to the chemical potential and
we replaced $\mu(\Gamma=0.5)$ by the chemical potential $- k_BT\ln (n_w/n)
+ \mu_{s}$
 of an ideal two-dimensional solution of ions in the surface layer of water
with two-dimensional
 concentration $n_w$. The term $\mu_{s}$ is the 
hydration free energy per ion at the surface and at $n \ll n_w$ does not depend on the
concentration of ions $n$. The first term of
Eq.~(\ref{cpot}) corresponds
to zero temperature WC and can be found directly from  Eq.~(\ref{energyion}).

We show below that when an counterion moves away from SCL, it leaves behind its
negatively charged correlation hole. We will also see
that the potential energy of attraction to this hole
becomes smaller than $k_BT$ at $x>l/4$.
On the other hand the selfconsistent potential of strongly screened surface is
so small that it
changes by $k_BT$ only at exponentially large length $\Lambda$, which is
defined below. Therefore the 
condition of equilibrium between SCL and the layer $l/4 \ll x \ll \Lambda$ is
\begin{equation}
\mu(n) = \mu(N),
\label{equilibrium}
\end{equation}
where $\mu(n)$ is given by Eq.~(\ref{sumchem}) and 
\begin{equation}
\mu(N) = - k_BT\ln N_w/N + \mu_{b}
\label{chem2}
\end{equation}
is the chemical potential of the bulk gas-like phase, $N_w$ is the bulk
concentration of
water and $\mu_{b}$ is the bulk hydration free energy which does not depend
on $N$. Using Eq.~(\ref{sumchem})
and Eq.~(\ref{chem2}) and solving Eq.~(\ref{equilibrium})
for $N$ we obtain at $l/4 \ll x \ll \Lambda$
\begin{equation}
N(0) = \frac{n}{w} \exp\left(-{{|\mu_{c}(n,T)|}\over{k_B T}}\right),
\label{bc0}
\end{equation}
where $w = (n_{w}/N_{w})\exp[(\mu_{b} - \mu_{s})/ k_BT)]$. Below we assume
for simplicity that
 $\mu_{b} = \mu_{s}$. In this case $w$ is the length 
of the order of size of the water molecule (for estimates we use $w = 0.3$ nm).

The notation $N(0)$ reflects the fact that this value plays the role of a
new boundary condition
 at $x\ll \Lambda$ for solution of PBE which as we will see has large
characteristic length $\Lambda$.
 Due to the large value of $|\mu_{c}(n,T)|$, the concentration $N(0)$ can be very small.
For example, at $Z=3$ and $\sigma = 1.0~e/$nm$^{-2}$, when $\Gamma = 6.4$
and according to Eq.~(\ref{cpot})
$|\mu_{c}(n,T)|/k_B T = 7.0$ we obtain $N(0) = 10^{24}$ m$^{-3}$ = 1.6
mM. Switching to
$Z=4$ we get $\Gamma = 9.9$, $|\mu_{c}(n,T)|/k_B T = 12.4$ and $N(0)= 5.5~\mu$M.
It is clear now that the $|\mu_c(n,T)|$ plays in our problem the role
similar to the work function
for thermal emission or to the free energy of chemisorption.
Thus we see that correlation effects in SCL provide nonspecific strong
binding of counterions to the macroion surface.  
Qualitative arguments for such binding can be found in Ref.~\onlinecite{Roland}

We would like to stress that such binding does not happen at $Z=1$. Indeed,
at $\sigma = 1.0~e/$nm$^{-2}$ one obtains from Eq.~(\ref{Gamma}) 
and Eq.~(\ref{cpot}) that $\Gamma = 1.2$ and $\mu_c(n,T)/k_BT = 0.4$.
Therefore the boundary condition Eq.~(\ref{bc0}) does not produce
nontrivial effects and standard solutions of PBE remain approximately valid.

Below we justify the role of the distance $l/4$ and give an idea how $N(x)$
evolves from $n/\lambda$ at $x\sim \lambda$ to $N(0)$ at $x=l/4$. Let us
move one ion of the
crystal along the $x$ axis. As we mentioned above, it leaves behind its
correlation hole. In the range
of distances $\lambda \ll x \ll R$, the correlation hole is approximately
the disc of charge
with radius $R$ (the Wigner-Seitz cell) and the ion is attracted to the surface
by its uniform
electric field $E = 2 \pi \sigma/ D$.
Therefore, if $\lambda$ were larger than $w$ we would get $N(x)=
(n/\lambda)\exp(-x/\lambda)$
at $x \ll R$. In the cases of our interest $\lambda < w $ and
$N = n/w$ at $x < w$, while at $w \ll x \ll R$ %
\begin{equation}
N(x) = \frac{n}{w}\exp(-x/\lambda).
\label{disc}
\end{equation}
At $ x \gg R$ the correlation hole radius grows and becomes of the order of
$x$. Indeed SCL can be considered as a good conductor
in the plane $(y, z)$.
It is known that a charge at distance $x$ from a metallic plane attracts an
opposite charge into
a disc with the radius $\sim x$ or, in other words, creates its point like
image on the other side of the plane at the distance $ 2x$ from the
original charge.
 The same thing happens to SCL. The removed ion repels other ions of SCL
 and creates a correlation hole in the form of the negative disc with the charge
 $-Ze$ and the radius $x$. The correlation hole attracts the removed ion
and decreases
 its potential energy by the Coulomb term
$U(x) = - Z^{2}e^{2}/ 4Dx$.
This effect provides the correction to
the activation energy of $N(x)$ :
\begin{equation}
N(x)=\frac{n}{w}\exp\left(-{{|\mu(n)| - Z^{2}e^{2}/4 Dx} \over{k_B
T}}\right)~~( x \gg R).
\label{Nx}
\end{equation}
The similar``image" correction to the work function of a metal is well-known
in the theory of thermal emission. The correction decreases with $x$ and at
$x =l/4$,
becomes equal to $k_B T$, so that $N(x)$ saturates at the value $N(0)$.
The dramatic difference between the exponential decay of Eqs.~(\ref{disc}),
(\ref{Nx}) and $1/x^2$ law of Eq.~(\ref{GC}) is obviously related to
correlation effects neglected in PBE.
Recall that it was assumed in the beginning of this paper that the charge
of the surface is almost totally compensated by SCL. Exponential decay of
$N(x)$ with $x$ confirms this assumption and at $\Gamma \gg 1$ makes this
theory self-consistent.

Consider now what happens with $N(x)$ at distances $x \gg l/4$. At such
distances,
correlations of the removed ion with its correlation hole in SCL are not
important
and correlation between ions of the gas phase are even weaker
because $N(x)$ is exponentially small. Therefore, one can return to PBE.
Solution of PBE with the boundary
condition (\ref{bc0}) and $N(\infty)= 0$ is similar to Eq.~(\ref{GC}):
\begin{equation}
N(x) = {1\over{2 \pi l}}{1\over {(\Lambda + x)^2}}~~~~~~(x \gg l/4),
\label{GCnew}
\end{equation}
where the new renormalized Gouy-Chapman length, $\Lambda$, is exponentially
large
\begin{equation}
\Lambda = (2\pi l N(0))^{-1/2}= \sqrt {\frac {w}{2 \pi n l}}
\exp\left({{|\mu_{c}(n,T)|}\over{2k_B T}}\right).
\label{Lambda}
\end{equation}
For example, at $\sigma = 1.0~e/$nm$^{-2}$ Eq.~(\ref{Lambda})
gives $\Lambda \simeq 0.8, 5.3, 68$ nm at $Z=2,3,4$. Comparing these lengths with 
$l/4 = 0.7, 1.6, 2.8$ nm respectively we see that 
$\Lambda \gg l/4$ for $Z \geq 3$, what justifies the use
of Eq.~(\ref{bc0}) as the boundary condition for the large distance
 solution of PBE. At $Z=2$, however, $\Lambda \sim l/4$ 
and our theory works only qualitatively. This is the reason why we do not 
compare our $N(x)$ quantitatively with numerical 
results~\cite{Roland}, obtained only for $Z=2$.

Using Eq.~(\ref{GCnew}) one finds that the total
surface charge density located at distances $x < l/4$
\begin{equation}
\sigma^* = - \sqrt {N(0) /(2\pi l_{B})} = - \sigma (\lambda/\Lambda).
\label{Sigma}
\end{equation}
The exponentially small $\sigma^*$ is a result of correlation effects,
which strongly bind counterions to the surface. Corrections to $\mu_{c}(n,T)$
and $N(0)$  related to such small $\sigma^*$ can be neglected.
Until now we talked about the case when $N(\infty)= 0$.
Let us now assume that there is a concentration $N(\infty)$ of $Z$:1 salt in
the bulk of solution, so that in the bulk $N(\infty)$ of counterions
is neutralized by  $N_{-}(\infty) = ZN(\infty)$ of monovalent coions.
 This adds the Debye-H\"{u}ckel screening radius
\begin{equation}
r_s = \left(4\pi l N(\infty) (1+Z^{-2})\right)^{-1/2} \label{screenrad}
\end{equation}
to the problem. If $N(\infty) \ll N(0)$ screening radius $r_s \gg \Lambda$
and the fact that $N(\infty)$ is finite
changes only the very tail of Eq.~(\ref{GCnew}) making decay of $N(x)$ at
$x \gg r_s $
exponential. At $x \ll r_s$ still $N(x) \gg ZN_{-}(x)$ and all previous
results are valid.
However, when $N(\infty)$ approaches $N(0)$, the solution changes
dramatically and $\sigma^*$ vanishes.
 Indeed when $N(\infty) = N(0)$ concentration
$N(x) = N(\infty)\exp(-{Ze\psi}/{k_BT}) = N(0)$ stays constant and
potential $\psi(x) = 0$ at $x > l/4$.
This means that the surface is completely neutralized at distances $0 <x <l/4$.

If $N(\infty) \gg N(0)$ negative charges dominate at $x \ll r_s$. Indeed in
the PBE approach,
$N(x) = N(\infty)\exp(-{Ze\psi}/{k_BT}), N_{-}(x) =N_{-}(\infty)\exp({e\psi}/{k_BT})$
and when concentration $N(x)$ decreases  $N_{-}(x)$ increases. One can
derive a boundary condition for $N_{-}(x)$ at $x=0$ from these equations
\begin{equation}
N_{-}(0) =ZN(\infty)[N(\infty)/N(0)]^{1/Z},
 \label{newbc}
\end{equation}
where $N(0)$ is given by Eq.~(\ref{bc0}).
Then the solution of PBE for $N_{-}(x)$ at $x \ll r_s$ has the form similar
to Eq.~(\ref{GCnew})
$N_{-}(x) = (2 \pi l_{B})^{-1}(\Lambda_{-} + x)^{-2}$,
where $\Lambda_{-} = (2\pi l_{B} N_{-}(0))^{-1/2}$.
To compensate the bulk negative
charge the positive surface charge density of SCL becomes larger than $\sigma$,
 so that the net surface charge density $\sigma^* > 0$. Similarly to
Eq.~(\ref{Sigma}), it is
\begin{equation}
\sigma^* = e~\sqrt {N_{-}(0) 2\pi l_{B}} = e/2\pi l_{B}\Lambda_{-}.
\label{Sigmaover}
\end{equation}
This phenomenon is called charge inversion and is, of course, impossible in
the framework of the
standard PBE. Technically, charge
inversion follows from the small value of $N(0)$ in Eq.~(\ref{bc0}).
Its physics is related to the strong binding of counterions at the charged
surface
due to formation of SCL. Remarkably,
when $\Gamma \gg 1$, this
phenomenon happens under the influence of very small concentration of salt.

Let us switch to the
screening of an infinite rigid cylinder with a radius $a$, a surface charge
density
$-\sigma$ or, in other words, with a linear charge density $\eta = -2\pi a
\sigma$.
 We assume that $\sigma$ is large enough so that
the surface of the cylinder
is covered by a two-dimensional SCL with $R < 2\pi a $ and with $\Gamma \gg 1$.
(This means that the following important inequality, $|\eta| \gg \eta_c = eZ/l$, is
satisfied).
Such a cylinder can be a first order approximation for the double helix DNA,
where $a =1~$nm, $\eta = 5.9~e/$nm, $\sigma =0.94~e/$nm$^{2}$, and at $Z=3$
length $R\simeq 1$~nm and $l=6.3~$nm.
Below we assume that, as in this example, $l \gg a$.

A screening atmosphere of a cylinder is described by concentration $N(r)$,
where $r$ is the distance from the cylinder axis.
For $|\eta| \gg \eta_c$ the solution of PBE is known~\cite{Zimm,Frank}
to confirm the main features of the Onsager-Manning~\cite{Manning}
picture of the counterion condensation. The screening charge $|\eta| -
\eta_c$ is located at the cylinder surface,
while the rest of the screening charge, $\eta_c$, at $N(\infty)=0$,
is spread in the bulk of the solution.
This means that the net charge density of the cylinder,
$\eta^*$, equals $- \eta_c$ and does not depend on $\eta$.
 At a finite $N(\infty)$ the charge density $\eta^*$ is screened only at
linear screening radius $r_s$.

Does SCL at the surface of the cylinder change these elegant statements? We
show below that
\begin{equation}
\eta^* = - \eta_c {{\ln [N(0)/N(\infty)]} \over\ln{[4/(\pi N(\infty)l^{3})]}}.
\label{etaapp}
\end{equation}
It is clear from Eq.~(\ref{etaapp}) that if two logarithms are close to
each other, i. e. if
$\ln (N(0)^{2}l^{3}/N(\infty)) \gg 1$
the Onsager-Manning theory is approximately correct and $\eta^*$ approaches
$- \eta_c$.
Note, however, that concentration $N(0)$ itself is exponentially small so that
values of $N(\infty)$ at which $\eta^*$ is close to
$- \eta_c$ are unrealistically small.
On the other hand, in disagreement with the Onsager-Manning theory one
obtains from Eq.~(\ref{etaapp})
that $|\eta^*| \ll \eta_c$ when concentration $N(\infty)$ of the salt is
still exponentially small,
namely $N(0)^{2}l^{3} \ll N(\infty) \ll N(0)$.
Moreover, according to Eq.~(\ref{etaapp}) $\eta^*$ vanishes at $N(\infty) =
N(0)$. This result is easy
to understand without calculations. Indeed, in this case $N(r) =
N(\infty)\exp(-Ze\psi(r)/k_BT) = N(0)$
stays constant and $\psi(r)=0$ at all $r > l/4$, so that all charge of the
polyelectrolyte is compensated
inside cylinder with $r=l/4$.
The difference from the Onsager-Manning theory becomes even more apparent at
$N(\infty) > N(0)$ when the density
$\eta^*$ becomes positive, once more demonstrating the charge inversion.
Note that the charge inversion takes
place still at exponentially small $N(\infty)$. One can show that
a positive $\eta^*$ continues to grow with growing $N(\infty)$
until it reaches critical density $e/l_B=\eta_c /Z$ at which Onsager-Manning
condensation of monovalent  coions begins.
Charge $\eta^*$ includes all counterions for which the binding energy exceeds
$k_BT$. Therefore, anomalous charge density of a
polyelectrolyte $\eta^*$ can be measured
in electrophoresis experiment. Finally, Eq.~(\ref{etaapp}) shows that
$\eta^*$  does depend on bare charge density
$\eta$ trough $n$ in Eq.~(\ref{bc0}). Thus, for screening by multivalent
ions at
$\Gamma \gg 1$ and at reasonably large $N(\infty)$ all
statements of Ref.~\onlinecite{Manning,Zimm} are qualitatively incorrect.

To derive  Eq.~(\ref{etaapp}), we focus at distances $l/4 < r < r_s$, where
the electrostatic potential of the linear charge density $\eta^*$ is not
screened and the boundary
 condition of Eq.~(\ref{bc0}) can be used to write
\begin{eqnarray}
N(r)&=&N(0)\exp\left(-{{Ze[\psi(r) -
\psi(l/4)]}\over{k_BT}}\right)\nonumber\\
&\simeq&N(0)\exp\left({{2\eta_a}\over{\eta_c}}\ln(4r/l)\right).
 \label{unscreenN}
\end{eqnarray}
At $ r = r_s$ concentration $N(r_s) \simeq N(\infty)$. Solving this
equation for $\eta^*$ we get
\begin{equation}
\eta^* = - \eta_c {{\ln [N(0)/N(\infty)]}\over{\ln(4r_s/l)^{2}}}.
 \label{etaapprs}
\end{equation}
Finally, using Eq.~(\ref{screenrad}), we arrive at
Eq.~(\ref{etaapp})

We are grateful to V. A. Bloomfield, M. M. Fogler, A. Yu. Grosberg, R.
Kjellander, S. Marcelja, R. Podgornik and I. Rouzina for valuable discussions.
This work was supported by RFSF (V. P.) and by NSF DMR-9616880.

\end{multicols}
\end{document}